\documentclass[11pt,english]{article}
\usepackage[T1]{fontenc}
\usepackage[latin9]{inputenc}

\makeatletter
\usepackage{hyperref}\pdfoutput=1
\makeatother

\usepackage{babel}

\makeatother

\usepackage{babel}

\begin{document}

\title{Vapour Cloud Dynamics Induced by Evaporation}

\author{Sam Dehaeck and Pierre Colinet \\
 \\
 \vspace{6pt}
 Transfers, Interfaces and Processes (TIPs), \\
 Universit\'e Libre de Bruxelles, CP 165/67, Brussels, Belgium}
\maketitle
\begin{abstract}
In this fluid dynamics video, the vapour cloud generated near evaporating
free liquid surfaces is visualised by Mach-Zehnder interferometry
(MZI). More precisely, for evaporating HFE-7100 (from 3M) and ambient
conditions, the vapour concentration field and its dynamical behavior
are observed in three simple experiments. Through these experiments,
it is evidenced that the high density of the vapour cloud (compared
to air) induces convective motions in the gas mixture, resulting in
deviations of the concentration field from a purely diffusional behavior.
\end{abstract}

\section{Introduction}

Vapour concentration levels can be measured by several techniques
such as infrared absorption \cite{vapourIrAbsorption} or planar laser
induced fluorescence PLIF \cite{vapourConcentrationPLIF}. These techniques
depend upon certain properties of the examined evaporating liquid,
such as suitable absorption bands or fluorescent behaviour and are
hence not applicable to every evaporating liquid.

Mach-Zehnder interferometry (MZI) is a very sensitive optical technique
enabling the measurement of dynamically evolving refractive index
fields in transparent media. This technique is already known to be
suitable for precise measurements of temperature or concentration
fields in liquids (e.g. \cite{dehaeckEvapCocktails,ColinetWylockDehaeckCO2Instab2008}).
However, for liquids with a large vapour pressure such as HFE-7100
(from 3M), changes in vapour concentration also lead to measureable
variations in optical path length, even for ambient conditions and
modest integration lengths of e.g. 2 mm. While the current technique
also depends upon a large dependence of the refractive index on the
vapour concentration and is hence not applicable to all liquids, it
is demonstrated that MZI nevertheless could complement infrared absorption
techniques and PLIF for the detailed study of vapour cloud dynamics.

\section{Step by Step Explanation}

The video available in the ancillary files starts with an explanation
of the MZI-setup. This is followed by a first experiment where a pending
drop of HFE-7100 is created and is hanging from a 2mm external diameter
needle in ambient air. The vapour cloud surrounding this drop is convected
downwards due to its higher density.

In the second experiment, a pending drop is created in ambient air
close to a solid substrate. The vapour cloud falls down and spreads
along the solid substrate. Liquid is continuously injected until a
liquid bridge is formed between the needle (1mm) and the solid substrate.
In this case, the isoconcentration lines in the vapour cloud adapt
themselves to this new liquid shape.

In the third and final experiment, a small rectangular cuvette (of
1cm by 1cm) is filled up to some height with HFE-7100 in ambient conditions.
Initially, a plastic cover prevents evaporation and a homogeneous
air-vapour layer sits on top of the liquid meniscus. When the cover
is suddenly removed, a fast evolution of the air-vapour layer is witnessed
and it transforms into a stagnant diffusive zone which occasionally
spills out some excess vapour over the edges of the cuvette. In addition,
B\'enard-like convection cells are visible in the liquid phase due to
the free surface cooling generated by evaporation.

\section{Acknowledgements}

The authors gratefully acknowledge financial support from: Cimex funded
by Esa and Belspo Prodex and Fonds de la Recherche Scientifique -
FNRS

\bibliographystyle{plain}

\end{document}